\documentclass[floatfix, preprint, showpacs, showkeys, preprintnumbers, nofootinbib, superscriptaddress]{revtex4-1}
\usepackage[utf8]{inputenc}
\usepackage[sort&compress]{natbib}
\usepackage{ulem}
\usepackage{overpic}
\usepackage{bm}
\usepackage{times}
\usepackage{amssymb,amsbsy,amsmath,amsfonts}
\usepackage{graphicx}
\usepackage{float}
\usepackage{color}
\usepackage{morefloats}
\usepackage{rotating}
\usepackage{srcltx}
\usepackage{slashed}
\usepackage{subfigure}
\usepackage{multirow}
\usepackage{verbatim}
\usepackage{hyperref}
\usepackage{tabularx}

\begin{document}

\title{$P_{c}(4457) \to P_{c}(4312) \pi/\gamma$ in the molecular picture  }

\author{Xi-Zhe Ling}
\affiliation{School of Physics, Beihang University, Beijing 102206, China}

\author{Jun-Xu Lu}
\affiliation{School of Space and Environment, Beihang University, Beijing 102206, China}
%\affiliation{School of Physics, Beihang University, Beijing 102206, China}

\author{Ming-Zhu Liu}\email{zhengmz11@buaa.edu.cn}
\affiliation{School of Space and Environment, Beihang University, Beijing 102206, China}
\affiliation{School of Physics, Beihang University, Beijing 102206, China}

\author{Li-Sheng Geng}\email{lisheng.geng@buaa.edu.cn}
\affiliation{School of Physics, Beihang University, Beijing 102206, China}
\affiliation{
Beijing Key Laboratory of Advanced Nuclear Materials and Physics,
Beihang University, Beijing 102206, China}
\affiliation{School of Physics and Microelectronics, Zhengzhou University, Zhengzhou, Henan 450001, China}
\affiliation{Beijing Advanced Innovation Center for Big Data-Based Precision Medicine, School of Medicine and Engineering, Beihang University, Beijing, 100191}

\begin{abstract}
The three pentaquark states, $P_{c}(4312)$, $P_{c}(4440)$, and $P_{c}(4457)$, discovered by the LHCb Collaboration in 2019, can be nicely arranged into a  multiplet of $\bar{D}^{(\ast)}\Sigma_{c}^{(\ast)}$ of seven molecules dictated by heavy quark spin symmetry. In this work we employ the effective Lagrangian approach to investigate the two decay modes of $P_{c}(4457)$,  $P_{c}(4457) \to P_{c}(4312) \pi$ and $P_{c}(4457) \to P_{c}(4312) \gamma$, via the triangle mechanism, assuming that $P_{c}(4457)$ and $P_{c}(4312)$ are $\bar{D}^{\ast}\Sigma_{c}$ and $\bar{D}\Sigma_{c}$ bound states  but  the spin of $P_{c}(4457)$ can be either 1/2 or 3/2. Our results show that the spin of $P_{c}(4457)$ can not be discriminated  through these two decay modes.      
The decay widths of $P_{c}(4457) \to P_{c}(4312) \pi$ and $P_{c}(4457) \to P_{c}(4312) \gamma$ are estimated to be of  order of 100 keV and 1 keV, respectively. The ratio of the partial decay widths of $P_{c}(4457) \to P_{c}(4312) \pi$ to $P_{c}(4457) \to P_{c}(4312) \gamma$ is similar to the ratio of $D^{\ast}\to D\pi$ to $D^{\ast}\to D\gamma$, which could be used to check the molecular nature of $P_{c}(4457)$ and $P_{c}(4312)$ if they can be observed in the future.  
  \end{abstract}
  
\maketitle

\section{Introduction: }

 The heavy quark spin symmetry (HQSS) dictates that the strong interaction is independent of the spin of the heavy quark in the limit of heavy quark masses~\cite{Isgur:1989vq,Isgur:1989ed}, which provides a natural explanation of the mass difference of $(D,D^{\ast})$ and $(B,B^{\ast})$, as well as those of their baryon counterparts. In the heavy quark mass limit, $D$ and $D^{\ast}$  as well as $\Sigma_{c}$ and $\Sigma_{c}^{\ast}$ belong to the same spin doublet, which has wide implications in charm physics~\cite{Yan:1992gz,Cheng:1992xi,Casalbuoni:1996pg}. Assuming that $D_{s1}(2460)$ and $D^*_{s0}(2317)$ are $D^{\ast}K$ and $DK$ molecules, they  can be regarded as a  HQSS doublet~\cite{Guo:2006rp,Altenbuchinger:2013vwa}. This molecule picture not only reconciles the quark model predictions~\cite{Godfrey:1985xj}  with the experimental measurements,  but also provides a self-consistent interpretation for the mass splitting of $D_{s1}(2460)$ and $D^*_{s0}(2317)$ in terms of that of $D$ and $D^{\ast}$. 
 
 It is interesting to note that in recent years, similar  multiplets of hadronic molecules seem to emerge in   other systems as well. In 2015, the LHCb Collaboration reported the observation of two resonant states, $P_{c}(4380)$ and $P_{c}(4450)$, in the $J/\psi p$ invariant mass distribution of the $\Lambda_{b}\rightarrow J/\psi p K$ decay~\cite{Aaij:2015tga}.  In 2019, they updated their analysis with a data set of  almost ten times bigger and found that the  $P_{c}(4450)$ state splits into two states, $P_{c}(4440)$ and $P_{c}(4457)$, and in addition a new narrow state $P_{c}(4312)$~\cite{Aaij:2019vzc} emerges just below the $\bar{D}\Sigma_c$ threshold. Their  masses and decay widths are
  \begin{eqnarray}
M_{P_{c}(4312)}&=&4311.9\pm 0.7^{+6.8}_{-0.6} ~\mbox{MeV}    \quad \quad   \Gamma_{P_{c}(4312)}=9.8 \pm 2.7^{+3.7}_{-4.5} ~\mbox{MeV},   \\ \nonumber
M_{P_{c}(4440)}&=&4440.3\pm 1.3^{+4.1}_{-4.7} ~\mbox{MeV}    \quad \quad   \Gamma_{P_{c}(4440)}=20.6 \pm 4.9^{+8.7}_{-10.1}~\mbox{MeV},
\\ \nonumber
M_{P_{c}(4457)}&=&4457.3\pm 0.6^{+4.1}_{-1.7} ~\mbox{MeV}    \quad \quad   \Gamma_{P_{c}(4457)}=6.4 \pm 2.0^{+5.7}_{-1.9} ~\mbox{MeV}.
\end{eqnarray}
 In our previous work we showed that these states can be interpreted  as $\bar{D}^{(\ast)}\Sigma_{c}$ hadronic molecules and predicted the existence of their HQSS partners in both the effective field theory (EFT) approach and the one boson exchange (OBE) model~\cite{Liu:2019tjn,Liu:2019stu}, where a complete multiplet of seven $\bar{D}^{(\ast)}\Sigma_{c}^{(\ast)}$ hadronic molecules dictated by HQSS presents itself, which has later been  corroborated by many studies~\cite{Xiao:2019aya,Yamaguchi:2019seo,Liu:2019zvb,Valderrama:2019chc,Du:2019pij}.
  Although at present
   the molecular interpretation is the most favored one~\cite{Xiao:2019aya,Xiao:2019mvs,Sakai:2019qph,Yamaguchi:2019seo,Liu:2019zvb,Valderrama:2019chc,Meng:2019ilv,Du:2019pij,Burns:2019iih,Wu:2019rog,Azizi:2020ogm,Phumphan:2021tta}, there exist other explanations, e.g.,
hadro-charmonium~\cite{Eides:2019tgv}, compact pentaquark states~\cite{Ali:2019npk,Mutuk:2019snd,Wang:2019got,Cheng:2019obk,Weng:2019ynv,Zhu:2019iwm,Pimikov:2019dyr,Ruangyoo:2021aoi}, 
virtual states~\cite{Fernandez-Ramirez:2019koa} or double triangle singularities~\cite{Nakamura:2021qvy}\footnote{ After $P_{c}(4380)$ and $P_{c}(4450)$ were discovered,  in Ref.~\cite{Guo:2015umn}, Guo et al.  pointed out that  $P_{c}(4450)$ might be caused by the triangle singularity mechanism because its mass coincides with the $\chi_{c1}p$ threshold. Later, Liu et al.  discussed more triangle diagrams which can produce peaks in the energy region where the pentaquark states were discovered~\cite{Liu:2015fea}. }.   See Refs.~\cite{Liu:2019zoy,Brambilla:2019esw,Guo:2019twa,Yang:2020atz,Dong:2021juy} for some latest reviews. As argued in Ref.~\cite{Pan:2019skd}, the most crucial, but still missing information to disentangle different interpretations is their spins.

To pin down the spins of $P_{c}(4440)$ and $P_{c}(4457)$  we can turn to  other systems which are related to the $\bar{D}^{(\ast)}\Sigma_{c}^{(\ast)}$ system via symmetries. 
In Refs.~\cite{Pan:2019skd,Pan:2020xek}, we extended the $\bar{D}^{(*)}\Sigma_c^{(*)}$ system to the $\Sigma_c^{(*)}\Xi_{cc}^{(\ast)}$ system via heavy antiquark diquark symmetry (HADS), and predicted the existence of a complete multiplet of ten triply charmed hadronic molecules. In particular, we pointed out that the mass splittings of $\Xi_{cc}\Sigma_c$ spin multiplets  are correlated with the spins of  $P_{c}(4440)$ and $P_{c}(4457)$, which, given the fact that the former can be much easily simulated on the lattice~\cite{Junnarkar:2019equ}, provides a  possibility to determine the spins of the later in a model independent way.
After the observation of a  hidden charm strange pentaquark, $P_{cs}(4459)$, with a statistical significance of $3.1$ $\sigma$  by the LHCb Collaboration~\cite{Aaij:2020gdg}, using the SU(3)-flavor symmetry we extended the  $\bar{D}^{(*)}\Sigma_c^{(*)}$ system to  the  $\bar{D}^{(*)}\Xi_c^{(\prime*)}$ system and predicted the existence of a multiplet of  $\bar{D}^{(*)}\Xi_c^{(\prime*)}$  hadronic molecules~\cite{Liu:2020hcv}. In addition, 
we found  that the existence of $\bar{D}^{(\ast)}\Xi_{c}$ molecules depends on the spins of $P_{c}(4440)$ and $P_{c}(4457)$ and the light quark configurations, and we pointed out that in one scenario the existence  $\bar{D}^{(\ast)}\Xi_{c}$ molecules can be used to determine the spins of  $P_{c}(4440)$ and $P_{c}(4457)$.

In Refs.~\cite{Lin:2019qiv,Xiao:2019mvs}, assuming that $P_{c}(4312)$, $P_{c}(4440)$, and $P_{c}(4457)$ are $\bar{D}^{(\ast)}\Sigma_{c}$ hadronic molecules, the authors adopted the effective Lagrangian approach to calculate the partial decay widths of three pentaquark states  to $J/\psi p$ via the triangle mechanism. However, the spins of  $P_{c}(4440)$ and $P_{c}(4457)$ cannot be  determined. Later,  the weak interaction of  $\Lambda_{b}\to \bar{D}_{s}^{(\ast)}\Sigma_{c}$ was studied via the $W$ boson emission and   the production rates of three pentaquark states  in the $\Lambda_{b}$ decay were calculated, from which the spins of   $P_{c}(4440)$ and $P_{c}(4457)$ cannot be determined either~\cite{Wu:2019rog}.   Following the same process, assuming that $P_{cs}(4459)$ is a $\bar{D}^{\ast}\Xi_{c}$ bound state with spin either 1/2 or 3/2, we calculated its production rate in the $\Xi_{b}$ decay, and we found that the present experimental data cannot fully  determine the $P_{cs}(4459)$ spin~\cite{Lu:2021irg}, although the decay width of a spin-3/2 $P_{cs}$ is larger than that of a spin-1/2 one by one order of magnitude with other things being equal.
\begin{figure}[!h]
\begin{center}
\begin{tabular}{c}
\begin{minipage}[t]{0.35\linewidth}
\begin{center}
\begin{overpic}[scale=.3]{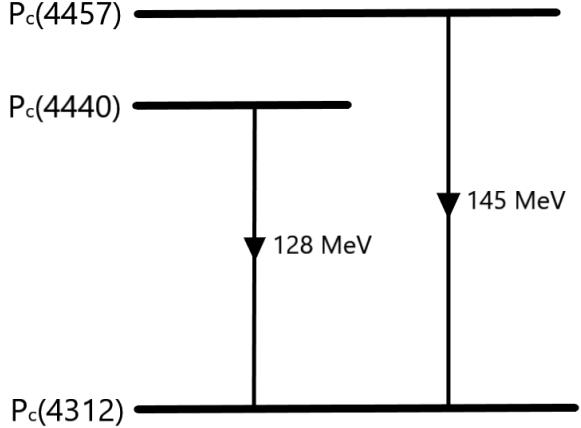}
		\put(90,8){}
\end{overpic}
\end{center}
\end{minipage}
\end{tabular}
\caption{ Mass splittings between $P_{c}(4457)$, $P_{c}(4440)$, and $P_{c}(4312)$.  }
\label{dec}
\end{center}
\end{figure}

In this work, we further investigate  the spin of   $P_{c}(4457)$ via its decay into   $P_{c}(4312)\pi$ and $P_{c}(4312)\gamma$. 
According to the  LHCb measurement~\cite{Aaij:2019vzc}, the mass splitting between $P_{c}(4440)$ and $P_{c}(4312)$ is 128 MeV, which is  less than the pion mass as shown in Fig.~\ref{dec}. Therefore, the decay of $P_{c}(4440) \to P_{c}(4312) \pi$ is forbidden  due to phase space.  The  mass splitting between $P_{c}(4457)$ and $P_{c}(4312)$ is 145 MeV, accordingly the   $P_{c}(4457)\to P_{c}(4312)\pi$ decay is allowed. Moreover, the radiative decays of  $P_{c}(4457)\to P_{c}(4312)\gamma$ and $P_{c}(4440)\to P_{c}(4312)\gamma$ are both allowed. We note that in Refs.~\cite{Faessler:2007gv,Faessler:2007us,Faessler:2008vc}, assuming that $D^*_{s0}(2317)$ and $D_{s1}(2460)$ are $DK$ and $D^{\ast}K$ bound states, Amand Faessler, et al  investigated the radiative and pionic decays  of $D_{s1}(2317)\to D^*_{s}\pi/\gamma$ and   $D_{s1}(2460)\to D^*_{s}\pi/\gamma$ together with their heavy quark spin partners, $B_{s0}^*(5725)$ and $ B_{s1}(5778)$  through the effective Lagrangian approach. Subsequently, Xiao, et al used the same approach to calculate the decay widths of
$D_{s1}(2460)\to D^*_{s1}(2317)\pi/\gamma$ via  the triangle diagrams~\cite{Xiao:2016hoa}. Based on the successful description of the strong and electromagnetic decays of $D_{s1}(2460) $ and $ D^*_{s1}(2317)$, the effective Lagrangian approach is extensively applied  to investigate decays of other exotic states in the molecular picture~\cite{Dong:2008gb,Dong:2009yp,Lu:2016nnt,Lin:2018kcc,Xiao:2020ltm}.  
In the present work,  assuming that $P_{c}(4457)$ and $P_{c}(4312)$ are  $\bar{D}^{(\ast)}\Sigma_{c}^{(\ast)}$  hadronic molecules, respectively,
we employ the effective Lagrangian approach to investigate the following two decay modes, $P_{c}(4457)\to P_{c}(4312)\pi$ and $P_{c}(4457)\to P_{c}(4312)\gamma$,  via the triangle mechanism.

  The manuscript is structured as follows. In Sec.~\ref{effective} we present the triangle diagram of the decays of $P_{c}(4457)\to P_{c}(4312)\pi$ and $P_{c}(4457)\to P_{c}(4312)\gamma$ as well as the relevant effective Lagrangians.
In Sec.~\ref{results} 
we provide the numerical results for the partial decay widths of $P_{c}(4457)\to P_{c}(4312)\pi$ and $P_{c}(4457)\to P_{c}(4312)\gamma$ and discuss their dependence on the cutoff of the regulator. 
Finally a brief summary  is given  in Sec.~\ref{summary}

\section{THEORETICAL FORMALISM}
\label{effective}
\begin{figure}[ttt]
\begin{center}
\begin{tabular}{cc}
\begin{minipage}[t]{0.5\linewidth}
\begin{center}
\begin{overpic}[scale=.7]{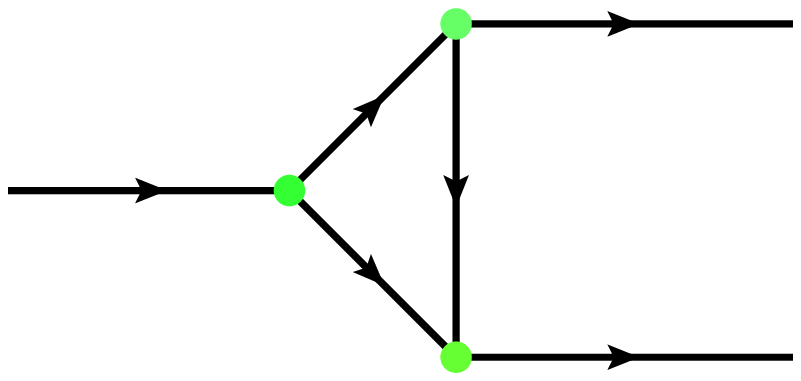}
		\put(70,6){$P_{c}(4312)$}
		
		\put(37,8){$\Sigma_{c}$}
		
		\put(37,38){$\bar{D}^{\ast}$}
		
		\put(2,26){$P_{c}(4457)$ }
		\put(75,38){$\pi$} \put(60,22){$\bar{D}$}
\end{overpic}
\end{center}
\end{minipage}
&
\begin{minipage}[t]{0.45\linewidth}
\begin{center}
\begin{overpic}[scale=.5]{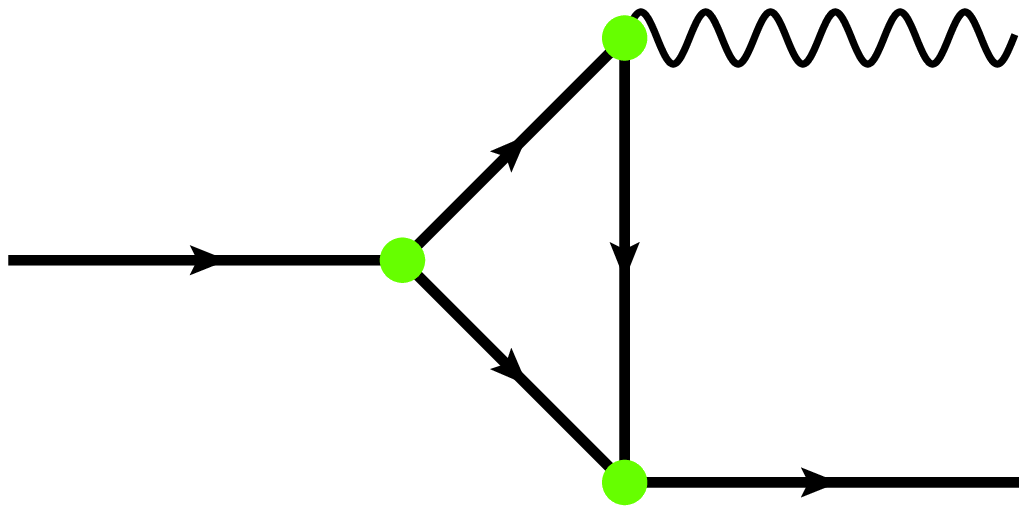}
		\put(70,6){$P_{c}(4312)$}
		
		\put(44,8){$\Sigma_{c}$}
		
		\put(44,38){$\bar{D}^{\ast}$}
		
		\put(7,27){$P_{c}(4457)$ }
		\put(75,38){$\gamma$} \put(63,22){$\bar{D}$}
\end{overpic}
\end{center}
\end{minipage}
\end{tabular}
\caption{ Triangle diagram of pionic and radiative  decay of $P_{c}(4457)$ to $P_{c}(4312)$ with the spin of $P_{c}(4457)$ being either 1/2 or 3/2.    }
\label{decay}
\end{center}
\end{figure}

First we explain how to  construct the triangle diagram to study the decay of  $P_{c}(4457)\to P_{c}(4312)\pi$.    Assuming $P_{c}(4457)$ and $P_{c}(4312)$ as $\bar{D}^{\ast}\Sigma_{c}$ and $\bar{D}\Sigma_{c}$ bound states,  the triangle mechanism for the $P_{c}(4457)\to P_{c}(4312)\pi$ decay is shown in Fig.~\ref{decay}, where  the $\bar{D}^{\ast}$ meson first  decays into $\bar{D}$ and $\pi$,  then the $\bar{D}$ and $\Sigma_{c}$ interaction dynamically generates   $P_{c}(4312)$. We note that the spin of $P_{c}(4457)$ is  not yet known experimentally. In the molecular picture,  its spin can be either 1/2 or 3/2 as an $S$-wave $\bar{D}^{\ast}\Sigma_{c}$ bound state. Therefore, one purpose of the present work is to check whether the pionic and radiative decay modes of $P_{c}(4457)$ can help us fix its spin.

\begin{figure}[ttt]
\centering
\begin{overpic}[scale=.8]{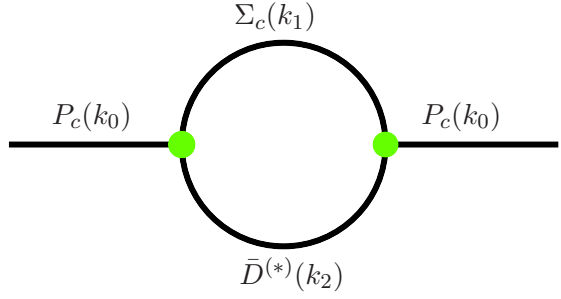}
\put(42,-6){$\bar{D}^{(\ast)}(k_{2})$} \put(41,41){$\Sigma_{c}(k_{1})$}
\put(8,23){$P_{c}(k_{0})$} \put(75,23){$P_{c}(k_{0})$}
\end{overpic}
\caption{Mass operators of  $P_{c}(4312)$ and $P_{c}(4457)$ as $\bar{D}\Sigma_{c}$ and $\bar{D}^{\ast}\Sigma_{c}$ bound states. }
\label{loop}
\end{figure}

Next we  explain how to calculate the triangle diagram of $P_{c}(4457)\to P_{c}(4312)\pi$ in the effective Lagrangian approach.    
The interactions between  $P_{c}(4312)$,  $P_{c}(4457)^{\frac{1}{2}^-}$, and  $P_{c}(4457)^{\frac{3}{2}^-}$, (denoted by $P_{c1}$, $P_{c2}$, and $P_{c3}$)  and their components can be described by the following Lagrangians~\cite{Xiao:2019mvs,Gutsche:2019mkg}, 
\begin{eqnarray}
\mathcal{L}_{P_{c1}\bar{D} \Sigma_{c}}&=&-i g_{P_{c1}\bar{D} \Sigma_{c}}  P_{c1}(x) \int dy \bar{D}(x+\omega_{\Sigma_{c}}y) \Sigma_{c}(x+\omega_{\bar{D}}y)\Phi(y^2) ,   \\ \nonumber
\mathcal{L}_{P_{c2}\bar{D}^{\ast} \Sigma_{c}}&=&g_{P_{c2}\bar{D}^{\ast} \Sigma_{c}} P_{c2}(x) \int dy \gamma^{\mu}\gamma_{5} \bar{D}^{\ast}_{\mu}(x+\omega_{\Sigma_{c}}y)\Sigma_{c}(x+\omega_{\bar{D}^{\ast}}y)\Phi(y^2),   \\ \nonumber
\mathcal{L}_{P_{c3}\bar{D}^{\ast} \Sigma_{c}}&=&g_{P_{c3}\bar{D}^{\ast} \Sigma_{c}} P_{c3}^{\mu}(x) \int dy \bar{D}^{\ast}_{\mu}(x+\omega_{\Sigma_{c}}y) \Sigma_{c}(x+\omega_{\bar{D}^{\ast}}y)\Phi(y^2),
\end{eqnarray}
where $\omega_{\Sigma_{c}}=\frac{m_{\Sigma_{c}}}{m_{\Sigma_{c}}+m_{\bar{D}^{(\ast)}}}$ and $\omega_{\bar{D}^{(\ast)}}=\frac{m_{\bar{D}^{(\ast)}}}{m_{\Sigma_{c}}+m_{\bar{D}^{(\ast)}}}$ are the kinematic parameters with $m_{\Sigma_{c}}$ and $m_{\bar{D}^{(\ast)}}$  the masses of involved particles, and  $g_{P_{c1}\bar{D} \Sigma_{c}}$, $g_{P_{c2}\bar{D}^{\ast} \Sigma_{c}}$, and $g_{P_{c3}\bar{D}^{\ast} \Sigma_{c}}$ are the couplings between the $\bar{D}^{(\ast)}\Sigma_{c}$ molecules and their corresponding components. The correlation function $\Phi(y^2)$ is introduced to reflect the distribution of the two components in a molecule, which also renders the Feynman diagrams
ultraviolet    finite.  Here we choose the Fourier transformation of the correlation
function in form of a Gaussian function 
\begin{eqnarray}
\Phi(p^2)=\mbox{Exp}(-\frac{p_{E}^{2}}{\Lambda^2}),
\end{eqnarray}
where $\Lambda$ is a size parameter, and $P_{E}$ is the Euclidean momentum.    The couplings of $g_{P_{c1}\bar{D} \Sigma_{c}}$, $g_{P_{c2}\bar{D}^{\ast} \Sigma_{c}}$, and $g_{P_{c3}\bar{D}^{\ast} \Sigma_{c}}$ can be estimated by reproducing the binding energies of the pentaquark states via the compositeness condition~\cite{Weinberg:1962hj,Salam:1962ap,Hayashi:1967bjx}.  The condition indicates that  the coupling constant can be determined from the fact that the renormalization constant of the wave function of a composite particle  should be zero. 
 For a spin-1/2 $\bar{D}^{(\ast)}\Sigma_{c}$ bound state, the compositeness condition is,
\begin{equation}
   Z_{P_{c}}=1-\frac{d \Sigma_{P_{c}}(k_{0})}{d
{k\!\!\!/}_0}|_{{{k\!\!\!/}_0=m_{P_{c}}}}=0,
\label{21}
\end{equation}
where $\Sigma_{P_{c}}(k_{0})$ is the self-energy of the composite particle with spin-1/2, as illustrated in Fig~\ref{loop}.  For a spin-3/2 $\bar{D}^{\ast}\Sigma_{c}$ bound state, the self energy  can be divided into  a transverse part and a longitudinal part, i.e.,
\begin{eqnarray}
\Sigma^{\mu\nu}=g^{\mu\nu}_{\bot} \Sigma^{T}(k_{0})+\frac{k_{0}^{\mu}k_{0}^{\nu}}{k_{0}^2}\Sigma^{L}(k_{0}).
\end{eqnarray}
The compositeness condition for a spin-3/2 composite particle can be estimated from its transverse part
\begin{equation}
   Z_{P_{c}}=1-\frac{d \Sigma_{P_{c}}^{T}(k_{0})}{d
{k\!\!\!/}_0}|_{{{k\!\!\!/}_0=m_{P_{c}}}}=0.
\label{23}
\end{equation}

With Eqs. (\ref{21}) and (\ref{23}), the $P_{c}(4312)$ and $P_{c}(4457)$ couplings to their components can be determined.  With the size parameter $\Lambda$=1 GeV, the corresponding couplings are determined as  $g_{P_{c1}\bar{D} \Sigma_{c}}=2.294$, $g_{P_{c2}\bar{D}^{\ast} \Sigma_{c}}=1.069$, $g_{P_{c3}\bar{D}^{\ast} \Sigma_{c}}=1.851$, consistent with Refs.~\cite{Sakai:2019qph,Xiao:2019aya,Wu:2019rog}.  

%{\color{red} shall we discuss the dependence on the size parameter.}

The Lagrangian describing  the $D^{\ast}$ decay into $D$ and $\pi$ is given by 
\begin{eqnarray}
\mathcal{L}_{D D^{\ast} \pi}&=& -i g_{D D^{\ast} \pi} (D \partial^{\mu}\pi D^{\ast\dag}_{\mu}-D_{\mu}^* \partial^{\mu} \pi  D^{\dag})
\end{eqnarray}
where  the coupling  $g_{D D^{\ast} \pi}$ is determined as $g_{D D^{\ast} \pi}=16.1$ by  reproducing the decay width of $D^{\ast+}\to D^{+} \pi^{0}$~\cite{Tanabashi:2018oca}. 

Utilizing all the relevant  Lagrangians, the amplitude of $P_{c}(4457)\to P_{c}(4312)\pi$ of Fig.~\ref{decay} 
 can be written as 
 \begin{eqnarray}
 \nonumber
 i\mathcal{M}_{1/2}&=&g_{P_{c1}\bar{D} \Sigma_{c}}g_{P_{c2}\bar{D}^{\ast} \Sigma_{c}} g_{D D^{\ast} \pi}\int \frac{d^{4}q}{(2\pi)^4}  \bar{u}_{p_{c1}}\frac{1}{{/\!\!\!k}_{1}-m_{\Sigma_{c}}}\frac{1}{q^{2}-m_{\bar{D}}^{2}}p_{2\alpha}\frac{-g^{\alpha\beta}+\frac{k_{2}^{\alpha}k_{2}^{\beta}}{m_{\bar{D}^{\ast}}^2}}{k_{2}^2-m_{\bar{D}^{\ast}}^2}\gamma_{\beta}\gamma_{5}u_{p_{c2}}F(q^2),  \\ 
 i\mathcal{M}_{3/2}&=&g_{P_{c1}\bar{D} \Sigma_{c}}g_{P_{c3}\bar{D}^{\ast} \Sigma_{c}} g_{D D^{\ast} \pi} \int \frac{d^{4}q}{(2\pi)^4}  \bar{u}_{p_{c1}}\frac{1}{{/\!\!\!k}_{1}-m_{\Sigma_{c}}}\frac{1}{q^{2}-m_{\bar{D}}^{2}}p_{2\alpha}\frac{-g^{\alpha\beta}+\frac{k_{2}^{\alpha}k_{2}^{\beta}}{m_{\bar{D}^{\ast}}^2}}{k_{2}^2-m_{\bar{D}^{\ast}}^2}u_{p_{c3}\beta}F(q^2), \nonumber
 \label{pi}
 \end{eqnarray}
 where $\mathcal{M}_{1/2}$ and $\mathcal{M}_{3/2}$ represent the amplitudes of $P_{c}(4457)\to P_{c}(4312)\pi$ for the cases of the $P_{c}(4457)$ spin being either 3/2 or 1/2,  
$p_{2}$, $k_{1}$, $k_{2}$, and $q$ denote the momenta of $\pi$, $\Sigma_{c}$, $\bar{D}^{\ast}$, and $\bar{D}$, and  $u_{P_{c2}}, u_{P_{c3}}, $ and $\bar{u}_{P_{c1}}$ are the initial and final spinors, respectively.   In addition, to eliminate the ultraviolet divergence of the above amplitudes, we  supplement the relevant vertices of exchanging a $\bar{D}$ meson with the following  monopolar form factor $F(q^2)$
 \begin{eqnarray}
 F(q^2)=\frac{\Lambda^2-m^2}{\Lambda^2-q^2}\frac{\Lambda_1^2-m^2}{\Lambda_1^2-q^2},
 \end{eqnarray}
 which also reflects the internal structure of hadrons, similar to  the OBE model~\cite{Liu:2018bkx}.       
Since there are two different types of vertices involving  $\bar{D}$ in Fig~\ref{decay}, we use different cutoff values, i.e., $\Lambda$ and $\Lambda_1$, for each type of vertices. Following Ref.~\cite{Yamaguchi:2011xb}, we assume that $\Lambda$ and $\Lambda_1$ are related and we study the following two scenarios to estimate the induced uncertainty, i.e., $\Lambda_1=0.9\Lambda$ and  $\Lambda_1=1.1\Lambda$.

 The radiative decay of $P_{c}(4457)[P_{c}(4440)]\to P_{c}(4312)+\gamma$ can also be investigated in the effective Lagrangian approach via the triangle mechanism shown in  Fig.~\ref{decay}. In comparison with with the $P_{c}(4457)\to P_{c}(4312)\pi$ decay, the vertex  $\bar{D}^{\ast}\to \bar{D}\pi$ in Fig.~\ref{decay} is replaced with $\bar{D}^{\ast}\to \bar{D}\gamma$. 
 The interaction between  charmed mesons and the photon is described by the following Lagrangian 
\begin{eqnarray}
\mathcal{L}_{D D^{\ast} \gamma}&=& \frac{g_{D D^{\ast} \gamma}}{4}e \varepsilon^{\mu\nu\alpha\beta}F_{\mu\nu}D^{\ast}_{\alpha\beta}D=e g_{D D^{\ast} \gamma}\varepsilon^{\mu\nu\alpha\beta} \partial_{\mu}A_{\nu}\partial_{\alpha}D^{\ast}_{\beta},
\end{eqnarray}
where  $F_{\mu\nu}=\partial_{\mu}A_{\nu}-\partial_{\nu}A_{\mu}$ and  $D^{\ast}_{\alpha\beta}=\partial_{\alpha}D^{\ast}_{\beta}-\partial_{\beta}D^{\ast}_{\alpha}$,  and the fine structure constant $\frac{e^2}{4\pi}=\frac{1}{137}$. The coupling  $g_{D D^{\ast} \gamma}$ is determined as 0.469 GeV$^{-1}$ by reproducing the decay width of $D^{\ast+}\to D^{+}\gamma$~\cite{Tanabashi:2018oca}.

The amplitudes of the triangle diagrams in Fig.~\ref{decay} 
 can be written as 
 \begin{eqnarray}
 \nonumber
 i\mathcal{M}_{1/2}&=&e g_{P_{c1}\bar{D} \Sigma_{c}}g_{P_{c2}\bar{D}^{\ast} \Sigma_{c}} g_{D D^{\ast} \gamma}\int \frac{d^{4}q}{(2\pi)^4}  \bar{u}_{p_{c1}}\frac{1}{{/\!\!\!k}_{1}-m_{\Sigma_{c}}}\frac{1}{q^{2}-m_{\bar{D}}^{2}}\varepsilon_{\mu\nu\sigma\alpha}p_{2}^{\mu}\varepsilon^{\nu}(p_{2})k_{2}^{\sigma}
 \\ \nonumber
 && \frac{-g^{\alpha\beta}+\frac{k_{2}^{\alpha}k_{2}^{\beta}}{m_{\bar{D}^{\ast}}^2}}{k_{2}^2-m_{\bar{D}^{\ast}}^2}\gamma_{\beta}\gamma_{5}u_{p_{c2}}F(q^2),  \\ \nonumber
 i\mathcal{M}_{3/2}&=&e g_{P_{c1}\bar{D} \Sigma_{c}}g_{P_{c3}\bar{D}^{\ast} \Sigma_{c}} g_{D D^{\ast} \gamma} \int \frac{d^{4}q}{(2\pi)^4}  \bar{u}_{p_{c1}}\frac{1}{{\slashed{k}}_{1}-m_{\Sigma_{c}}}\frac{1}{q^{2}-m_{\bar{D}}^{2}}\varepsilon_{\mu\nu\sigma\alpha}p_{2}^{\mu}\varepsilon^{\nu}(p_{2})k_{2}^{\sigma}
 \\  \nonumber
 &&  \frac{-g^{\alpha\beta}+\frac{k_{2}^{\alpha}k_{2}^{\beta}}{m_{\bar{D}^{\ast}}^2}}{k_{2}^2-m_{\bar{D}^{\ast}}^2}u_{p_{c3}\beta}F(q^2),\nonumber
 \label{photon}
 \end{eqnarray}
 where $\mathcal{M}_{1/2}$ and $\mathcal{M}_{3/2}$ are for the initial state having spin 1/2 and 3/2, respectively,    $p_{2}$ $k_{1}$, $k_{2}$, and $q$ denote the momentum of photon,  $\Sigma_{c}$, $\bar{D}^{\ast}$, and $\bar{D}$, respectively, and  $u_{P_{c2}}, u_{P_{c3}} $, and $\bar{u}_{P_{c1}}$ represent the initial and final spinors. In the mechanism shown in Fig.~\ref{decay}, the radiative decay of $P_{c}(4457)\to P_{c}(4312)+\gamma$ occurs via its component  $\bar{D}^{\ast}$.
 \begin{figure}[!h]
\begin{center}
\begin{tabular}{c}
\begin{minipage}[t]{0.45\linewidth}
\begin{center}
\begin{overpic}[scale=.6]{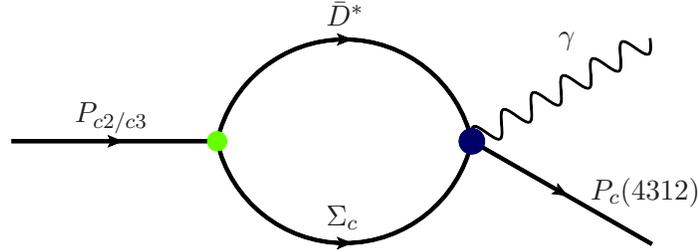}
		\put(90,8){$P_{c}(4312)$}
		
		\put(49,4){$\Sigma_{c}$}
		
		\put(49,35){$\bar{D}^{\ast}$}
		
		\put(10,20){$P_{c2/c3}$ }
		\put(85,32){$\gamma$} 
\end{overpic}
\end{center}
\end{minipage}
\end{tabular}
\caption{  Contact diagram of the radiative decay of $P_{c}(4457)$ to $P_{c}(4312)$ with the spin of $P_{c}(4457)$ being either 1/2 or 3/2.    }
\label{decayr}
\end{center}
\end{figure}

 For the description of radiative decays,  gauge invariance is important. If the spins of the initial and final states are both 1/2,  the electromagnetic current can be written as 
 \begin{eqnarray}
 \langle \bar{u}(p_{1}) | J^{\mu} | u(k_{0}) \rangle =  \bar{u}(p_{1}) (g_{1} \gamma^{\mu}+g_{2}\frac{p_{1}^{\mu}{p\!\!\!/}_{2}}{p_{1}\cdot p_{2}}) u(k_{0}),
 \end{eqnarray}
 where $J^{\mu}$ should satisfy the identity relationship $p_{2\mu}J^{\mu}=0$. The loop integral of $\mathcal{M}_{1/2}$ can be parameterized as 
 \begin{eqnarray}
 \mathcal{M}^{Tri}_{1/2}=\varepsilon_{\mu}(p_{2})  \bar{u}(p_{1}) (g_{1}^{Tri1/2} \gamma^{\mu}+g_{2}^{Tri1/2}\frac{p_{1}^{\mu}{p\!\!\!/}_{2}}{p_{1}\cdot p_{2}}) u(k_{0}).
 \end{eqnarray}
 If  $g_{1}^{Tri1/2}$ is not equal to $g_{2}^{Tri1/2}$, we have to add the contact term shown in Fig.~\ref{decayr} to satisfy gauge invariance. Our numerical results show that  $g_{1}^{Tri1/2}$ is equal to $g_{2}^{Tri1/2}$. Thus the contact term is not necessary in our study.    
 
 If the spin of  the initial state is 3/2 and  that of the final state is 1/2,  the electromagnetic current can be written as 
 \begin{eqnarray}
 \langle \bar{u}(p_{1})    |  J^{\mu\nu} | u_{\nu}(k_{0}) \rangle= \bar{u}(p_{1})   ( g_{1}\gamma^{\mu} p_{2}^{\nu}- g_{2}{p\!\!\!/}_{2} g^{\mu\nu} ) u_{\nu}(k_{0}), 
 \end{eqnarray}
where  the tensor current should satisfy the relationship  $p_{2\mu}k_{0\nu}J^{\mu\nu}$=0. The amplitude of  $\mathcal{M}_{3/2}$ can be written as 
 \begin{eqnarray}
 \mathcal{M}^{Tri}_{3/2}=\varepsilon_{\mu}(p_{2})  \bar{u}(p_{1})  ( g_{1}^{Tri3/2}\gamma^{\mu} p_{2}^{\nu}- g_{2}^{Tri3/2}{p\!\!\!/}_{2} g^{\mu\nu} ) u_{\nu}(k_{0}).
 \end{eqnarray}  
The term $g_{1}^{Tri3/2}$  is found equal to $g_{2}^{Tri3/2}$, which implies that gauge invariance is satisfied.

 With the amplitudes of pionic and radiative decays of $P_{c}(4457)$    to $P_{c}(4312)$ determined, one can obtain 
 the corresponding partial decay widths  as
 \begin{eqnarray}
\Gamma=\frac{1}{2J+1}\frac{1}{8\pi}\frac{|\vec{p}|}{m_{P_{c}}^2}\bar{|\mathcal{M}|}^{2},
\end{eqnarray}
where $J$ is the total angular momentum of the initial state $P_{c}(4457)$, the overline indicates the sum over the polarization vectors of final states, and $|\vec{p}|$ is the momentum of either final state in the rest frame of  $P_{c}(4457)$.

\section{ NUMERICAL RESULTS AND DISCUSSIONS}
\label{results}

Before presenting the numerical results for the pionic and radiative partial decay widths,  we  discuss the cutoff dependence of our results. As  explained above, we have introduced a monopolar form factor for the meson exchange vertices.  The cutoff reflects the fact that hadrons are not pointlike particles and its value is not known a priori. In the OBE model the cutoff can be fixed by reproducing the binding energies of some molecular candidates. In our previous works~\cite{Liu:2019stu,Liu:2019zvb,Liu:2018bkx}, based on the molecular picture where the deuteron,  $X(3872)$, and $P_{c}(4312)$ are nucleon-nucleon, $\bar{D}D^{\ast}$, and $\bar{D}\Sigma_{c}$ bound states, we fixed the corresponding  cutoff of the OBE model as 0.86, 1.01, and 1.12 GeV, respectively, which can also be described by  an  empirical formula 
$\Lambda=m_{E}+\alpha\Lambda_{QCD}$~\cite{Xiao:2018kfx},  where $m_{E}$ is the most massive particle of all allowed exchange particles,  $\Lambda_{QCD}\sim 200-300$ MeV is the scale parameter of Quantum Chromodynamics (QCD), and $\alpha$ is a dismensionless parameter.
In general, $\alpha$ is  taken to be unity, and the corresponding formula is $\Lambda=m_{E}+\Lambda_{QCD}$. If we apply this formula to the present work, the cutoff is estimated to $\Lambda=2.1$ GeV.  In the following, to study the  dependence of the results on  the cutoff, we vary the cutoff from 2.1 to 2.6 GeV.

\begin{figure}[!h]
\centering
\subfigure[]
{
\centering 
\begin{overpic}[scale=.35]{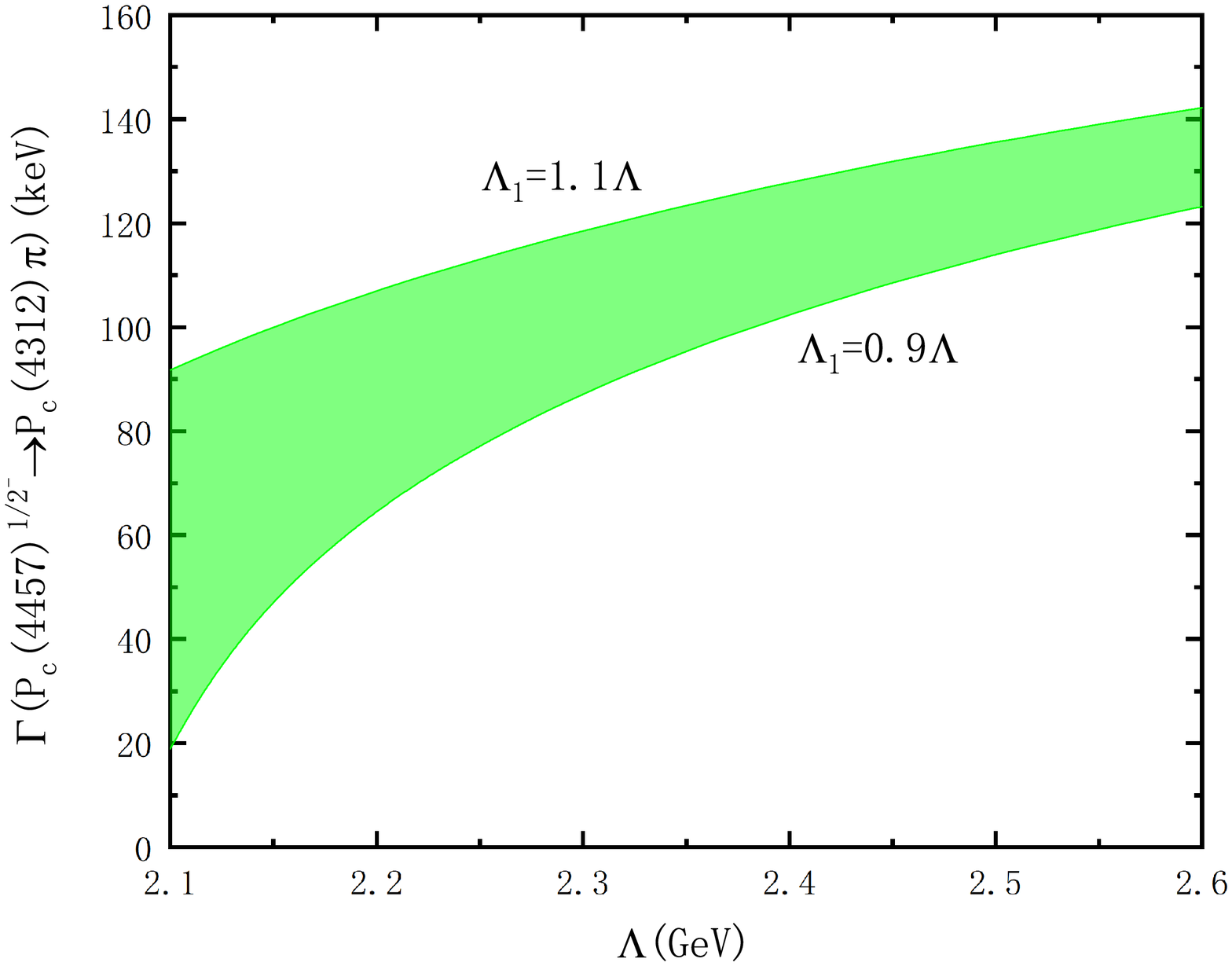}
\end{overpic}
}
\subfigure[]
{
\centering 
\begin{overpic}[scale=.35]{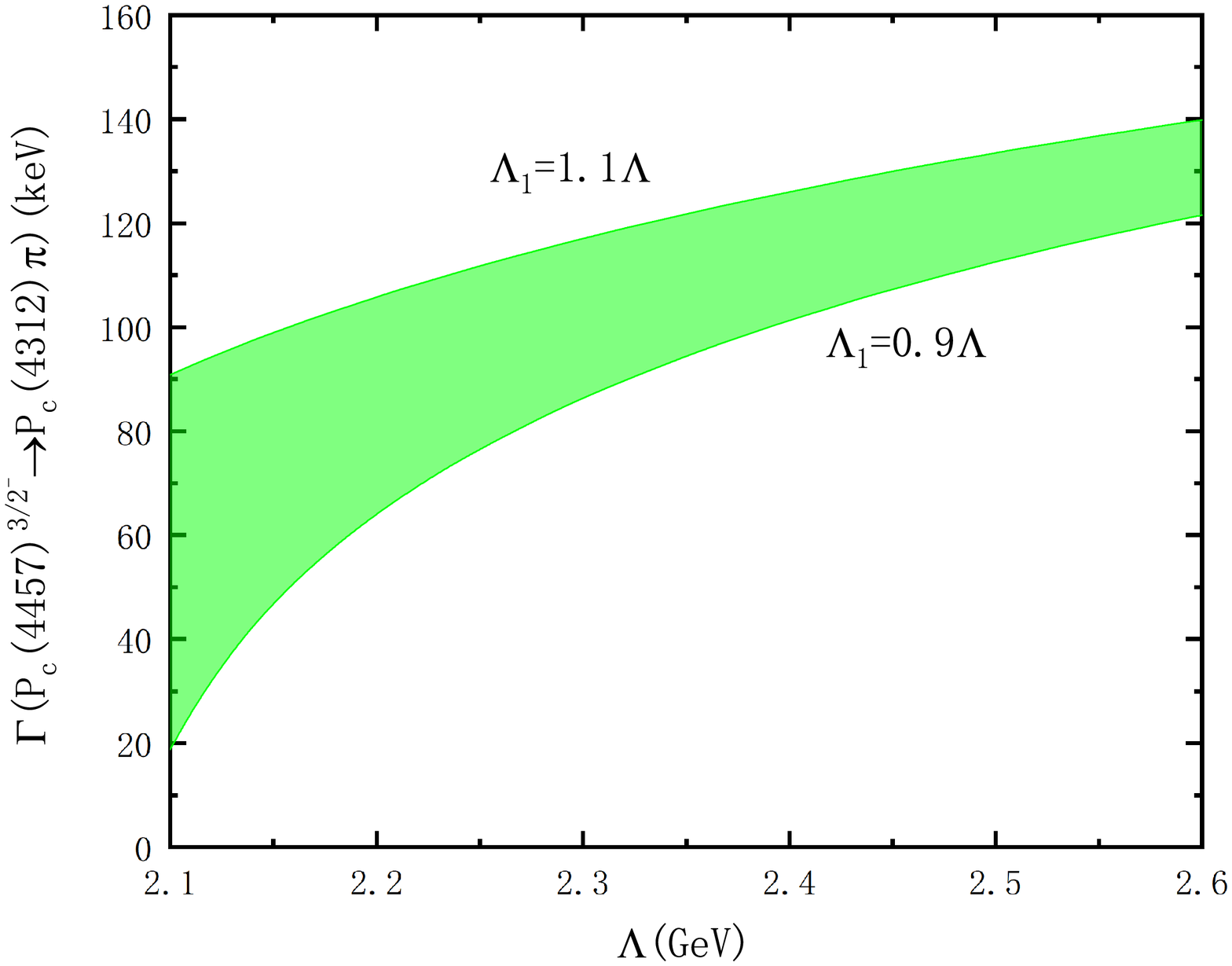}
\end{overpic}
}
\caption{Decay widths of $P_{c}(4457)^{1/2}\to P_{c}(4312)\pi$ (a) and $P_{c}(4457)^{3/2}\to P_{c}(4312)\pi$ (b) as a function of the cutoff.   }
\label{results1}
\end{figure}

\begin{figure}[!h]
\centering
\begin{overpic}[scale=.4]{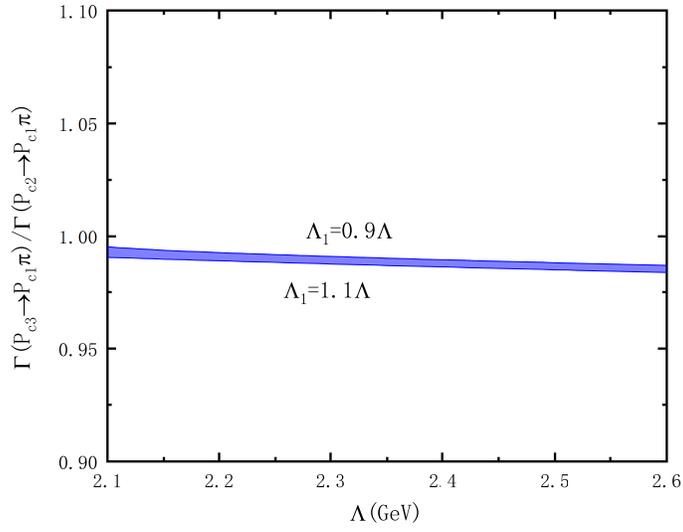}
\end{overpic}
\caption{Ratio of the partial decay widths of $P_{c}(4457)^{3/2}\to P_{c}(4312)\pi$ to  $P_{c}(4457)^{1/2}\to P_{c}(4312)\pi$ as a function of the cutoff.}
\label{results2}
\end{figure}

In Fig.~\ref{results1} we show the  decay width of $P_{c}(4457)\to P_{c}(4312)\pi$ as a function of the cutoff, where Fig.~(1a) is  for spin-1/2, and Fig.~(1b) for spin-3/2.
As the cutoff varies from 2.1 to 2.6 GeV, the  decay width of $P_{c}(4457)^{\frac{1}{2}^-}\to P_{c}(4312)\pi$  changes  from 
 91.7 keV to 142.2 keV with $\Lambda_1=1.1\Lambda$ and from  18.8 keV to 123.2 keV with $\Lambda_1=0.9\Lambda$, while  the  decay width of $P_{c}(4457)^{\frac{3}{2}^-}\to P_{c}(4312)\pi$  changes  from 
 90.9 keV to 139.8 keV with $\Lambda_1=1.1\Lambda$ and from  18.7 keV to 121.5 keV with $\Lambda_1=0.9\Lambda$. The  decay widths of  $P_{c}(4457)\to P_{c}(4312)\pi$ for spin 1/2 and 3/2 are close to each other, with the latter a bit larger, which can  be  explicitly seen from  the ratio of the decay width of a spin-3/2 $P_{c}(4457)$  to that of a spin-1/2 $P_c(4457)$ shown in Fig.~\ref{results2}.  Our results  show that the spin of $P_{c}(4457)$  can not be discriminated from the $P_{c}(4457)\to P_{c}(4312)\pi$ decay. The decay width of  $P_{c}(4457)\to P_{c}(4312)\pi$ is  predicted to be of order of 100 keV in the the molecular picture. In comparison with the total decay width of $P_{c}(4457)$, the branching ratio of  $P_{c}(4457)\to P_{c}(4312)\pi$ is 1.5$\%$.  With more data accumulated, the LHCb Collaboration should be able to observe  
the decay of  $P_{c}(4457)\to P_{c}(4312)\pi$.

\begin{figure}[ttt]
\centering
\subfigure[]
{
\centering 
\begin{overpic}[scale=.35]{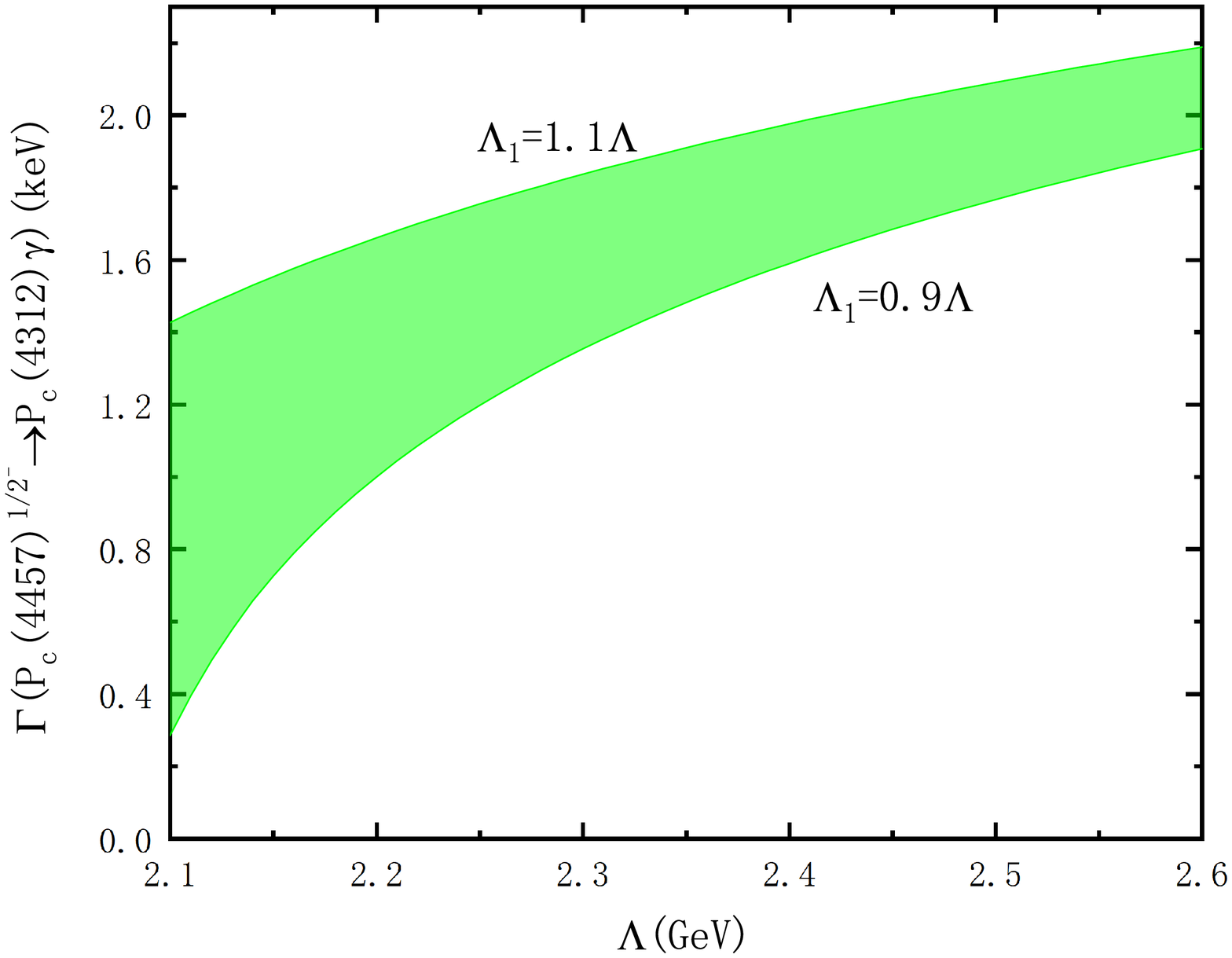}
\end{overpic}
}
\subfigure[]
{
\centering 
\begin{overpic}[scale=.35]{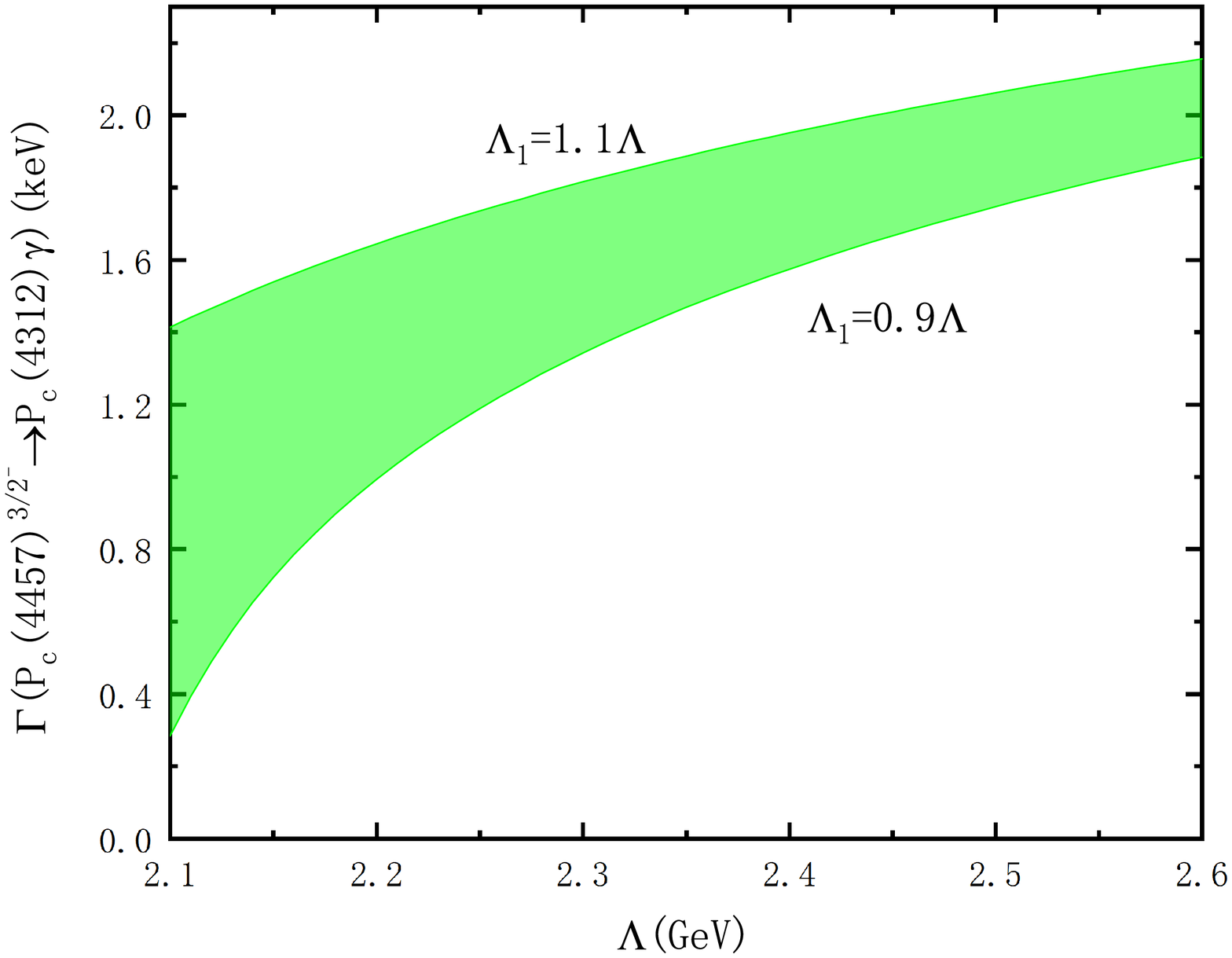}
\end{overpic}
}
\caption{Decay width of $P_{c}(4457)^{1/2}\to P_{c}(4312)\gamma$ (a) and $P_{c}(4457)^{3/2}\to P_{c}(4312)\gamma$ (b) as a function of the cutoff.   }
\label{results3}
\end{figure}

\begin{figure}[ttt]
\centering
\begin{overpic}[scale=.4]{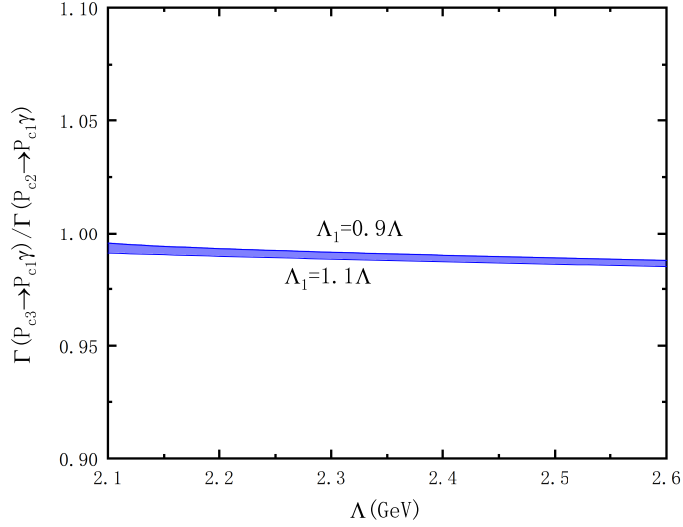}
\end{overpic}
\caption{Ratio of the decay width of  $P_{c}(4457)^{3/2}\to P_{c}(4312)\gamma$ to $P_{c}(4457)^{1/2}\to P_{c}(4312)\gamma$  as a function of the cutoff. }
\label{results4}
\end{figure}

For  the radiative decay, both $P_{c}(4457)$ and $P_{c}(4440)$ can decay into $P_{c}(4312)$. The spin of $P_{c}(4440)$  can be either 1/2 or 3/2 as a $\bar{D}^{\ast}\Sigma_{c}$ bound state, thus yielding similar results as the $P_{c}(4457)$ for the radiative decay, except for the small difference originating from the slightly different phase space. Therefore,  we refrain from explicitly presenting the numerical  results of  $P_{c}(4440)\to P_{c}(4312)\gamma$.
The  partial decay width of  $P_{c}(4457)\to P_{c}(4312)\gamma$ as a function of the cutoff is shown in Fig.~\ref{results3}, where the left and right figures are  for the spin-1/2 and spin-3/2 assignments for the initial state $P_{c}(4457)$, respectively. One can see that the decay widths are close to each other.  In Fig. ~\ref{results4} the ratio of the partial decay width of $P_{c}(4457)^{3/2}\to P_{c}(4312)\gamma$  to $P_{c}(4457)^{1/2}\to P_{c}(4312)\gamma$ is presented, which shows that the partial decay width for spin-3/2 is a bit larger than that for spin 1/2.  Similar to the pionic decay mode, the radiative decay mode cannot be employed to distinguish the two spin assignments for $P_{c}(4457)$. As the cutoff varies from 2.1 to 2.6 GeV, the decay width of $P_{c}(4457)^{\frac{1}{2}^-}\to P_{c}(4312)\gamma$ changes  from  1.4 keV to 2.2 keV with $\Lambda_1=1.1\Lambda$ and from  0.3 keV to 1.9 keV with $\Lambda_1=0.9\Lambda$, while the decay width of $P_{c}(4457)^{\frac{3}{2}^-}\to P_{c}(4312)\gamma$ changes  from  1.4 keV to 2.2 keV with $\Lambda_1=1.1\Lambda$ and from  0.3 keV to 1.9 keV with $\Lambda_1=0.9\Lambda$ . Compared with the total decay width of $P_{c}(4457)$, the corresponding branching ratio is about $0.02\%$.     

\begin{figure}[!h]
\centering
\subfigure[]
{
\centering 
\begin{overpic}[scale=.34]{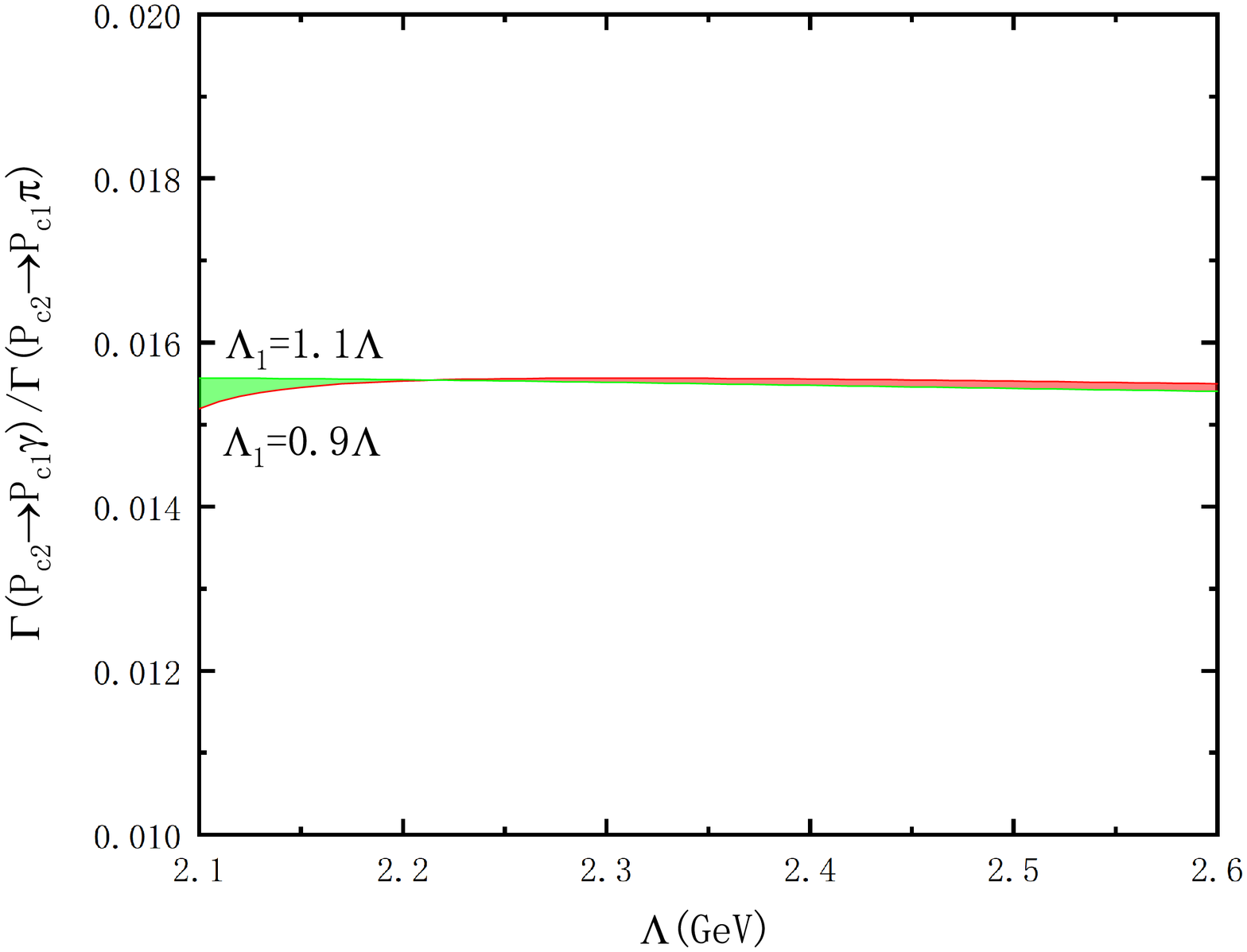}
\end{overpic}
}
\subfigure[]
{
\centering 
\begin{overpic}[scale=.34]{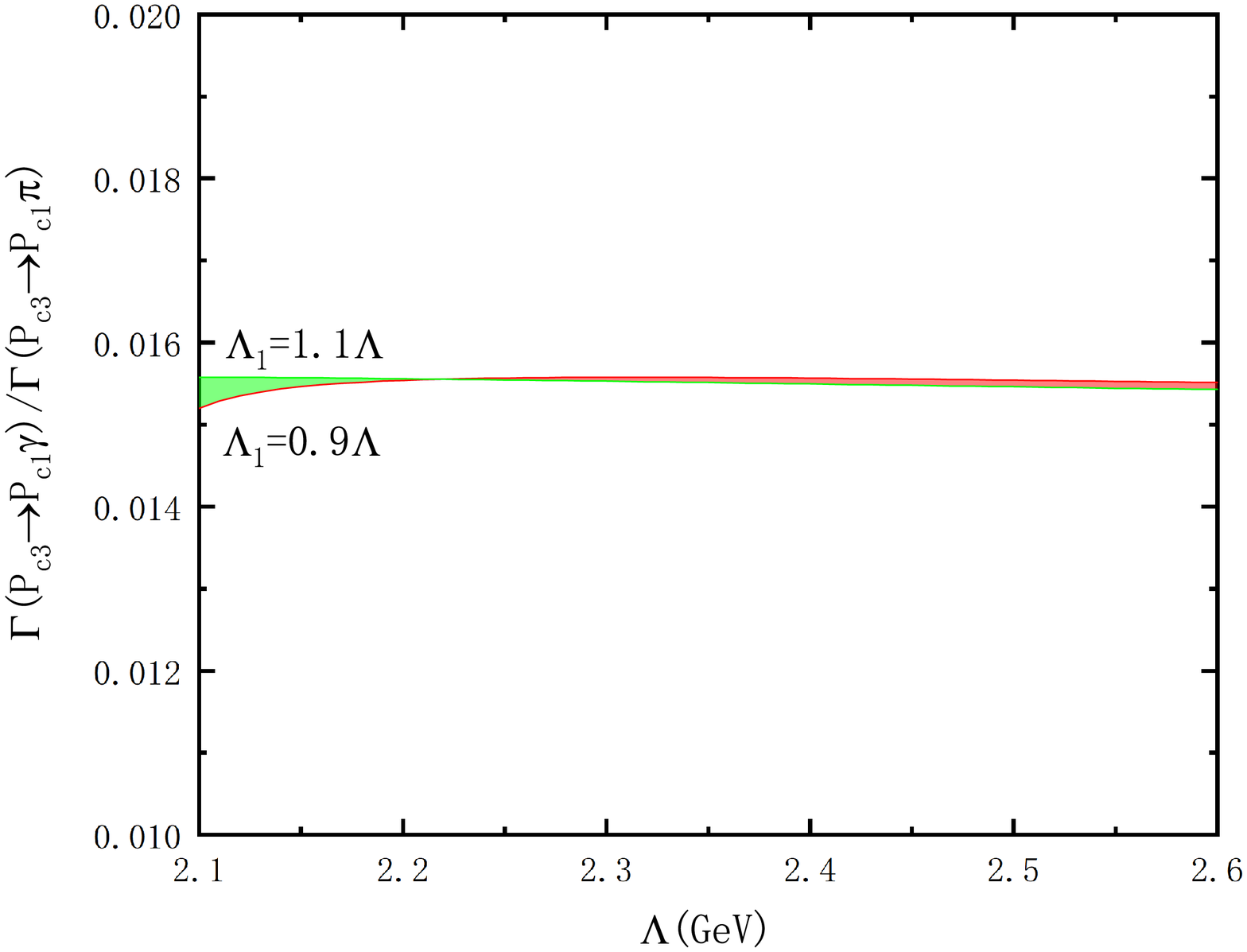}
\end{overpic}
}
\caption{Ratio of the radiative decay width and the pionic decay width of $P_{c}(4457)$ as a function of the cutoff.   }
\label{results5}
\end{figure}

If the spin of $P_{c}(4457)$ is 1/2,  the radiative decay $P_{c}(4457)\to P_{c}(4312)\gamma$ can be parameterized by two terms, electric-charge($E0$) and magnetic dipole($M1$), while only the $M1$ term
contributes to its radiative decay width. If the spin of  $P_{c}(4457)$ is 3/2, the radiative decay $P_{c}(4457)\to P_{c}(4312)\gamma$ is parameterzied by three terms, magnetic dipole($M1$),  electric quadrupole($E2$), and electric charge quadrupole($C1$), while only the $M1$ and $E2$ terms contribute to the radiative decay width. According to Refs.~\cite{Butler:1993ht,Alexandrou:2003ea,Aliev:2014bma,Bahtiyar:2015sga},  the  $E2$ contribution relative to that of $M1$ for a general baryon transition  $B(3/2)\to B(1/2) \gamma $ is of the order of $1\%$. Therefore, we can safely neglect the $E2$  contribution to  the radiative decay of $P_{c}(4457)\to P_{c}(4312)\gamma$.\footnote{In Ref.~\cite{Li:2021ryu}, the suppression of  the $E2$ term for  the $P_{c}(4457)\to P_{c}(4312)\gamma$ decay  is estimated to be $[(m_{P_{c}(4457)}-m_{P_{c}(4312)})/m_{P_{c}(4457)}]^{2}=0.1\%$.  }

Finally, the  ratio of the partial decay width of $P_{c}(4457)\to P_{c}(4312)\gamma$ to $P_{c}(4457)\to P_{c}(4312)\pi$ is given in Fig.~\ref{results5}, where the left and right figures are for spin-1/2 and spin-3/2 assignments, respectively.    The ratio is about 1.5$\%$ in the two cases,  in agreement with the ratio of the decay width of  $D^{\ast}\to D\gamma$ to $D^{\ast}\to D \pi$~\cite{Tanabashi:2018oca}, 
\begin{eqnarray}
\frac{Br(P_{c}(4457)\to P_{c}(4312)\gamma)}{Br(P_{c}(4457)\to P_{c}(4312)\pi)} \approx\frac{Br( D^{\ast}\to D\gamma)}{Br( D^{\ast}\to D\pi)}\sim  1.6 \%,
\end{eqnarray}
which illustrates that the radiative and pionic decays of charmed mesons may play an important role in describing the radiative and pionic  decays of $P_{c}(4457)$. In other words, $P_{c}(4457)$  and $P_{c}(4312)$ contain large molecular components, which reminds us of the fact that the  mass splitting of $D^*_{s0}(2317)$ and $D_{s1}(2460)$ can be easily understood in the   $DK$ and $D^{\ast}K$ molecular picture. Therefore,  if the ratio of  $P_{c}(4457)\to P_{c}(4312)\gamma$ to $P_{c}(4457)\to P_{c}(4312)\pi$ is observed in future experiments, it will help us to either confirm or repute the molecular nature of  $P_{c}(4457)$ and $P_{c}(4312)$.

\section{summary}
\label{summary}

In this work, assuming $P_{c}(4457)$ and $P_{c}(4312)$ as   $\bar{D}^{(\ast)}\Sigma_{c}^{(\ast)}$ hadronic molecules,   we  employ  the effective Lagrangian approach to investigate the decays of  $P_{c}(4457)\to P_{c}(4312)\pi$ and $P_{c}(4457)\to P_{c}(4312)\gamma$ via  the triangle mechanism.
With the assignment for the spin of  $P_{c}(4457)$  as either 1/2 and 3/2 we found the decay width of   $P_{c}(4457)\to P_{c}(4312)\pi$ is at
the order of 100 keV, and the decay width of $P_{c}(4457)\to P_{c}(4312)\gamma$ is at the order of   1.5 keV. In comparison with the total decay width of $P_{c}(4457)$, we obtained the  branching ratios of $P_{c}(4457)\to P_{c}(4312)\pi$ and $P_{c}(4457)\to P_{c}(4312)\gamma$, i.e.,  1.5$\%$ and 0.02$\%$, respectively. With a larger statistics, it is likely that the LHCb Collaboration could observe the  $P_{c}(4457)\to P_{c}(4312)\pi$ decay. The ratio of $P_{c}(4457)\to P_{c}(4312)\gamma$ to $P_{c}(4457)\to P_{c}(4312)\pi$ is 1.5$\%$, consistent with the ratio of $D^{\ast}\to D\gamma$ 
to $D^{\ast}\to D\pi$, which can be easily understood in the molecular picture where $P_{c}(4457)$ and $P_{c}(4312)$  are $\bar{D}^{\ast}\Sigma_{c}$ and $\bar{D}\Sigma_{c}$ molecules, respectively.   
In addition we found that     
the spin of  $P_{c}(4457)$ can not be discriminated through these two decay modes,    $P_{c}(4457)\to P_{c}(4312)\pi$ and  $P_{c}(4457)\to P_{c}(4312)\gamma$.

One should note that Voloshin argued that if the $\bar{D}^{\ast}\Sigma_{c}$ bound state has spin $3/2$, it can not decay into $\eta_{c}p$ in  $S$-wave.  On the other hand, if it has spin $1/2$,  the branching ratio  $\Gamma[P_{c}(J=1/2)\to \eta_{c}p]/\Gamma[P_{c}(J=1/2)\to J/ \psi p]$  is $3/25$ in the limit of  heavy quark masses. As a result, the observation of the decay mode to $\eta_c p$ will be helpful to determine the spins of $P_{c}(4440)$ and $P_{c}(4457)$~\cite{Voloshin:2019aut}.  Utilyzing heavy quark spin symmetry and kinematic effects,  Sakai et al.  also argued that the $\eta_{c}p$ channel can shed light on the spin of the $\bar{D}^{\ast}\Sigma_{c}$ bound state~\cite{Sakai:2019qph}. 
In Ref.~\cite{Du:2021fmf}, Du et al. used an effective field theory which takes into account inelastic $\bar{D}^{\ast}\Lambda_{c}$ and $\eta_{c}p$ channels  to fit the LHCb  data,  and they argued that the spins  of $P_{c}(4440)$ and $P_{c}(4457)$ can be determined by studying the line shapes of $\bar{D}\Sigma_{c}^{(\ast)}$ and $\eta_{c}p$. In addition,  in a number of works~\cite{Lin:2019qiv,Wang:2019spc,Xiao:2020frg,Chen:2020pac}, it was argued that the decays of the $\bar{D}^{(\ast)}\Sigma_{c}$ molecules into  $\eta_{c}p$ and $\bar{D}\Lambda_{c}$   could help  discriminate the spins of $P_{c}(4440)$ and $P_{c}(4457)$ and check the molecular nature of these pentaquark states. 

\section{Acknowledgments}
This work is partly supported by the National Natural Science Foundation of China under Grants Nos.11735003, 11975041, and 11961141004, and the fundamental Research Funds
for the Central Universities.

\bibliography{xiccsigmac}

\end{document}